\documentclass[twocolumn,aps,prl]{revtex4}
\usepackage[T1]{fontenc}
\usepackage[latin9]{inputenc}
\usepackage{amsmath}
\usepackage{amssymb}
\usepackage{graphicx}
\usepackage{esint}
\usepackage{color}
\usepackage{mathtools}
\usepackage{float}
\usepackage{placeins}

\DeclarePairedDelimiterX\braket[2]{\langle}{\rangle}{#1 \delimsize\vert #2}

\makeatletter

\definecolor{Green}{RGB}{0,204,102}
\definecolor{Purple}{RGB}{102,0,255}
\definecolor{Blue}{RGB}{51,153,255}
\definecolor{Red}{RGB}{201,010,010}

\makeatother

\begin{document}

\title{Dynamics of elliptical vortices in a trapped quantum fluid}

\author{Chuanzhou Zhu$^1$, Mark E. Siemens$^2$, Mark T. Lusk$^1$}

\affiliation{$^1$Department of Physics, Colorado School of Mines, Golden, CO 80401, USA$^*$\\
	$^2$Department of Physics and Astronomy, University of Denver, Denver, CO 80208, USA}
\email{mlusk@mines.edu, Mark.Siemens@du.edu}

\date{\today}

\begin{abstract}
The nonequilibrium dynamics of vortices in 2D quantum fluids can be predicted by accounting for the way in which vortex ellipticity is coupled to the gradient in background fluid density. In the absence of nonlinear interactions, a harmonically trapped fluid can be analyzed analytically to show that single vortices will move in an elliptic trajectory that has the same orientation and aspect ratio as the vortex projection itself. This allows the vortex ellipticity to be estimated through observation of its trajectory. A combination of analysis and numerical simulation is then used to show that nonlinear interactions cause the vortex orientation to precess, and that the rate of vortex precession is once again mimicked by a precession of the elliptical trajectory. Both vortex ellipticity and rate of precession can therefore be inferred by observing its motion in a trap. An ability to anticipate and control local vortex structure and vortex trajectory is expected to prove useful in designing few-vortex systems in which ellipticity is a ubiquitous, as-yet-unharnessed feature. 
\end{abstract}
\maketitle

\section{Introduction}

Quantum vortices \cite{Dark_Soliton} with orbital angular momentum have been widely studied in Bose-Einstein
condensates (BEC) \cite{Review_BEC} and optical fluids \cite{Review_Optics}.
Although these vortices typically have circular cross-sections in equilibrium, their non-equilibrium counterparts tend to be elliptical in both superfluid \cite{tilted_vortex_BEC2}
and optical~\cite{tilted_vortex_optics1, Andersen2021} settings. In fact, non-circular shapes are expected whenever two or more vortices interact, as in the generation/annihilation
of vortex-antivortex pairs~\cite{Antivortex_BEC1, Antivortex_BEC2, Andersen2021}, the merging of co-rotating vortices~\cite{corotating1, corotating2, corotating3}, and the braiding of vortex pairs~\cite{Braiding}. Such vortices do not move with the underlying fluid, as in incompressible flows, nor can their trajectories be anticipated by accounting for the influence of density gradients~\cite{Nilsen2006}.  
Ellipticity introduces two additional degrees of freedom that couple to the gradients in the \emph{background quantum state}, and a vortex velocity relation has recently been derived that correctly incorporates this and applies it to predict the motion of optical vortices in linear media~\cite{Andersen2021}.

In this paper, we elucidate the relationship between trap strength, nonlinear interaction, and the motion of an isolated, non-circular vortex in a quantum fluid. We find that a circular harmonic trap causes an elliptic vortex to move in an elliptical trajectory that, surprisingly, has the same orientation and aspect ratio as the vortex projection. Nonlinear interactions, on the other hand, induce a precession in the vortex ellipticity that is mimicked by an analogous precession in the vortex trajectory. We further find that a strong nonlinear interaction induces an oscillation of the aspect ratio of the elliptical vortex.

\section{Linear Quantum Fluid}

\begin{figure}[t]
	\begin{center}
		\includegraphics[width=0.85\linewidth]{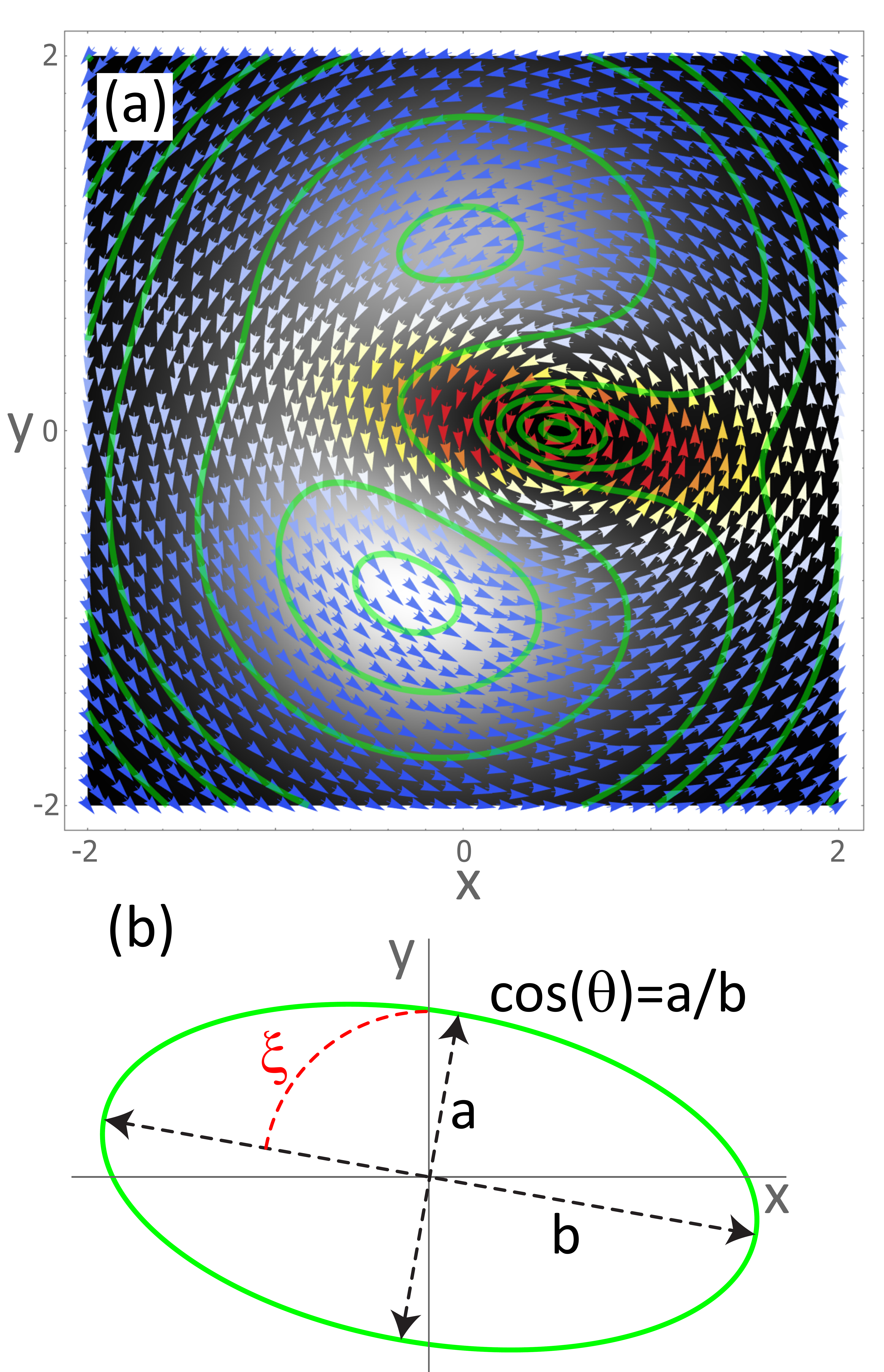}
	\end{center}
	\caption{ \emph{Vortex on the Shoulder of a Gaussian Density Distribution}. An unbounded, linear, compressible 2D fluid is given a Gaussian density distribution and an elliptical vortex $(\xi=80^{\circ}, \theta=60^{\circ})$, about the point $\{x,y\}=\{0.5,0\}$. (a) Regions of low density are dark, higher densities are light, and selected density contours are shown as solid curves (green online). The fluid velocity is depicted as colored arrows, with higher velocities in red and lower velocities in blue. (b) A single contour in the neighborhood of the vortex is an ellipse with orientation $\xi$ and an aspect ratio of $cos(\theta)$.} 
	\label{Fig1}
\end{figure}

The motion of non-circular vortices in linear media serves as a useful point of reference for understanding how nonlinear effects change their dynamics. In this simpler setting, the evolving quantum state, $\psi\left(x,y,t\right)$,
is assumed to be governed by the 2D Schr\"odinger equation,
\begin{equation}\label{linearSE}
	i\partial_{t}\psi = -\frac{1}{2}\left(\partial_{xx}+\partial_{yy}\right) + {\cal V}\psi ,
\end{equation}
where $\cal V$ represents potentials such as a harmonic trap. The initial fluid state has a Gaussian density profile implanted with a non-circular vortex offset from the center of the fluid by $x_{0}$:
\begin{equation}
	\psi_{0}\left(x,y\right)=Ne^{-\frac{1}{2}\left(x^{2}+y^{2}\right)}\left[\left(x-x_{0}\right)a+yb\right].\label{eq:psi0}
\end{equation}
Here 
\begin{equation}
	a=-\cos\xi+i\cos\theta\sin\xi, \,
	b=-\sin\xi-i\cos\theta\cos\xi
\end{equation}
and $N$ is a normalization factor. A hydrodynamic interpretation of this state~\cite{Madelung1927} is that the fluid density is $\left|\psi_{0}\right|^{2}$ and the fluid velocity is $\nabla \rm{Arg}(\psi_{0})$. Both are plotted in Fig.~\ref{Fig1}. The vortex core in our 2D quantum fluid is characterized by a point with phase singularity and zero fluid density in the 2D plane. The vortex shape is determined by the streamlines around this vortex core, and its amplitude is a linear function of radial position.
%
%
%

\subsection{Freely Expanding Linear Quantum Fluid} 

Suppose that the fluid is unbounded so that ${\cal V}=0$ in Eq. \ref{linearSE}. Then convolution of the initial state of Eq. \ref{eq:psi0} with the 2D Schr\"odinger Green function~\cite{FourierFresnelPaper} immediately describes the time evolution:
\begin{eqnarray}\label{psi_Gaussian_shoulder}
	&&\psi(x,y,t) =(\cos\xi(-x+x_0+\imath x_0 t - \imath y \cos\theta) \nonumber \\
	&&- y \sin\xi + \imath x \cos\theta\sin\xi + x_0(-\imath+t)\cos\theta_i\sin\xi)\\
	&&\times \frac{1}{(-\imath + t)^2}\sqrt{\frac{2}{\pi}}e^{\imath\frac{x^2+y^2}{2(-\imath+t)}}. \nonumber
\end{eqnarray}
%

\begin{figure}[t]
	\begin{center}
		\includegraphics[width=0.85\linewidth]{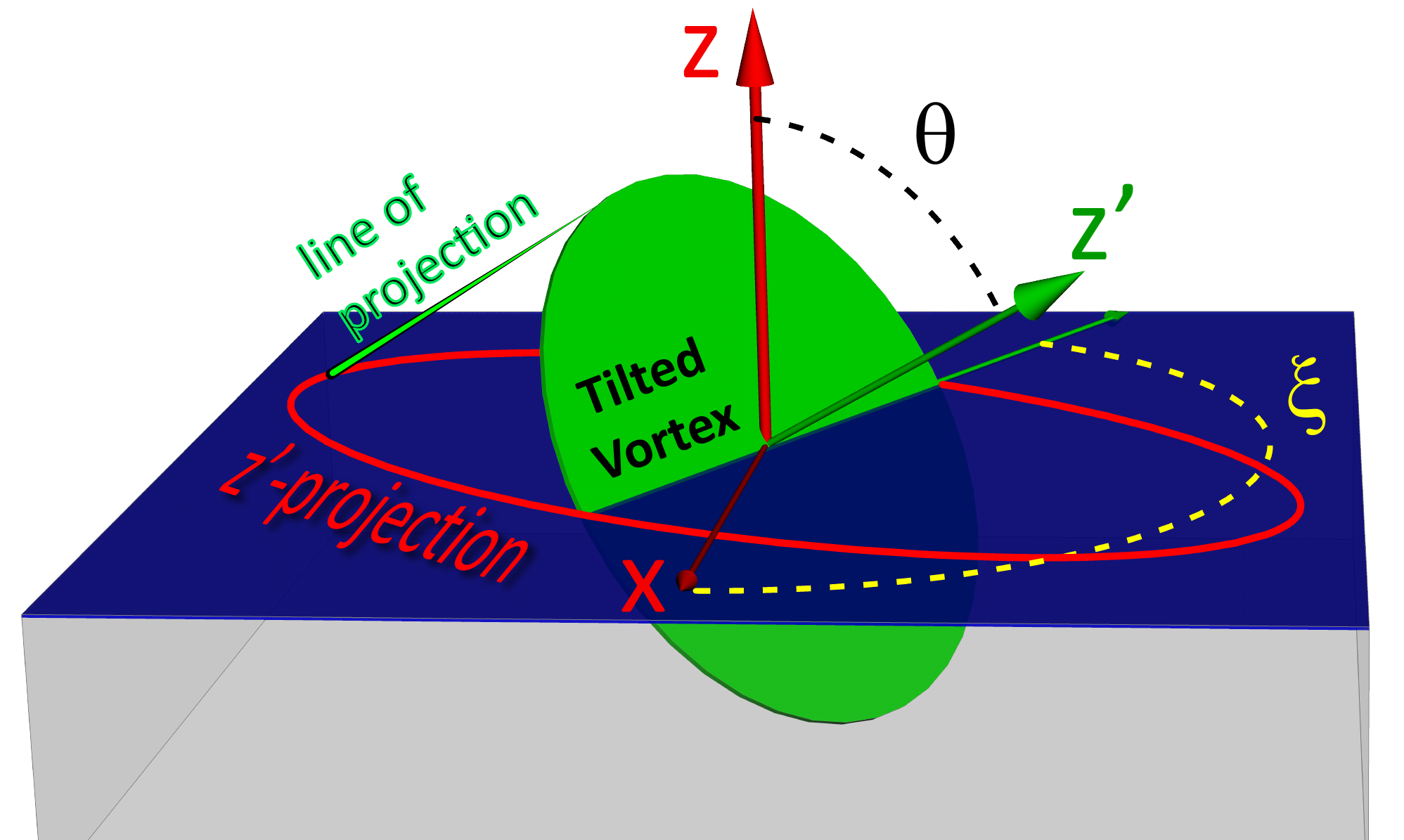}
	\end{center}
	\caption{ \emph{Tilted Vortex Perspective}. A 2D elliptical vortex in the transverse plane can be viewed as the projection, along the tilt axis, of a 3D vortex. The ellipticity can then be described by an azimuthal orientation angle, $\xi$, and a polar lean, $\theta$.} 
	\label{Fig2}
\end{figure}

We seek the trajectory of the vortex, and this can be obtained if its velocity is quantified as a function of time. Towards this end, it will prove both mathematically convenient and physically insightful to develop a formalism for quantifying the vortex ellipticity by visualizing vortices as the projection of a virtual construct, a circular vortex with an axis of symmetry described by tilt angles $\xi$ and $\theta$ as shown in Fig. \ref{Fig2}. In the northern hemisphere, polar leans, $\theta$, can be achieved from $0^{\circ}$ to $86^{\circ}$  in an optical fluid~\cite{Andersen2021} and up to at least $45^{\circ}$ in a BEC~\cite{tilted_vortex_BEC2}. A comparable range is associated with polar leans in the southern hemisphere.

%

A variational argument~\cite{Andersen2021} can then be used to show that the vortex velocity is described by 
\begin{equation}\label{v_final}
	\vec{v} = \vec{v}^{\varphi} + \vec{v}^{\rho},
\end{equation}
where 
\begin{equation}\label{v_phi_v_rho}
\vec{v}^{\varphi} = \nabla_\perp \varphi_{bg}, \quad
\vec{v}^{\rho} =  - \boldsymbol\Lambda \sigma_{0} \nabla_\perp \log\rho_{bg}.
\end{equation}
%
Here $\varphi_{bg}$ and $\rho_{bg}$ are the phase and magnitude of the background field, $\psi_{bg}=\rho_{bg}e^{i\varphi_{bg}}$, obtained by dividing out the contribution from the vortex itself. Their gradients each contribute to the vortex velocity denoted by $\vec{v}^{\varphi}$ and $\vec{v}^{\rho}$, respectively.
The Pauli-like operator, $\sigma_{0}$, in the 2D x-y plane is defined as
\begin{equation}
\sigma_{0}=\left(\begin{array}{cc}
0 & -1\\
1 & 0
\end{array}\right).
\end{equation}
The influence of vortex tilt is captured by the 2D tensor, $\boldsymbol\Lambda$. In the coordinate frame of Fig.~\ref{Fig2}, its elements are
\begin{align}\label{Lambda}
	\boldsymbol\Lambda_{xx} = \cos\theta \cos^2\xi + \sec\theta\sin^2\xi, \\
	\boldsymbol\Lambda_{yy} = \sec\theta \cos^2\xi + \cos\theta\sin^2\xi, \\
	\boldsymbol\Lambda_{xy} = \boldsymbol\Lambda_{yx} = \frac{-\sin(2\xi)\sin^2\theta}{2\cos\theta}.
\end{align}
%
This shows that the tilt of a vortex affects its motion because it is coupled to the local gradient in the background density, a result valid for both linear and nonlinear quantum fluids. 

In the limit of a circular, untilted vortex with $\theta = 0$, the $\boldsymbol\Lambda$ term in Eq. \ref{v_final} becomes a 2D unit tensor, which is consistent with the vortex velocity relations derived in earlier works \cite{Nilsen2006,Groszek2018}. The vortex velocity relation for a circular vortex has also been derived for the Ginzburg-Landau model~\cite{Ola1997}. 

\begin{figure}[t]
	\begin{center}
		\includegraphics[width=0.95\linewidth]{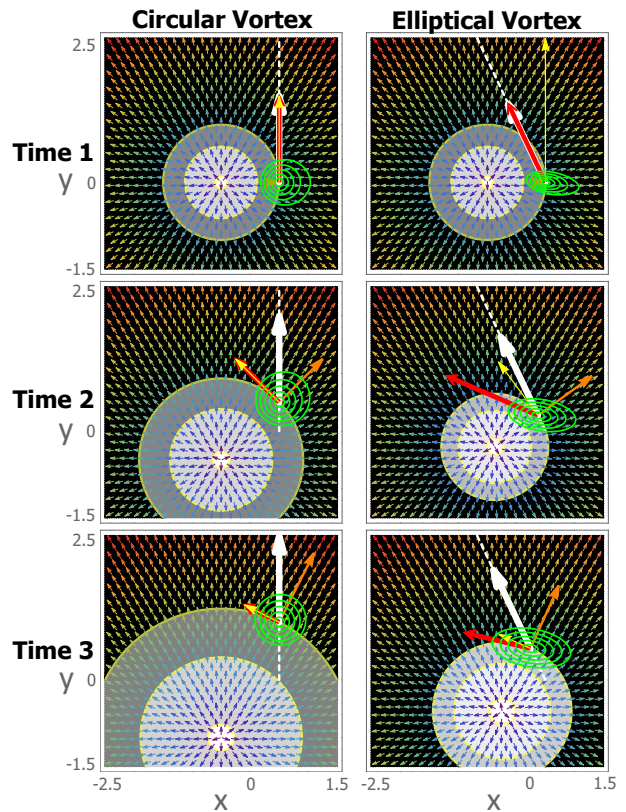}
	\end{center}
	\caption{ \emph{Vortex Velocity Components}.  An unbounded, linear, compressible 2D fluid is given a Gaussian density distribution and a vortex about the point {x,y} = {0.5,0}. The left column shows three subsequent time slices for a circular vortex, while the right column are associated with an elliptical vortex. Regions of low(high) \emph{background} density are dark(light), and selected background density contours are shown in yellow.  Green contours indicate the local density of the vortex part of the fluid. \emph{Background} fluid velocity is depicted as colored arrows. The vortex velocity is shown in white with its trajectory indicated with a dashed white line. For each panel, the orange (medium thickness) arrow shows the background fluid velocity at the vortex, the yellow (thin) arrow is the contribution to vortex velocity from the local density gradient with tilt disregarded, and the red (thick) arrow gives the actual contribution from the density gradient with tilt accounted for. The yellow arrows are tangent to background density contours. The actual vortex velocity (white) is the sum of red and orange arrows.} 
	\label{Fig3}
\end{figure}

The vortex velocity expression, Eq. \ref{v_final}, is applied to the evolving system state, Eq. \ref{psi_Gaussian_shoulder}, to obtain a prediction for direction and speed of vortex motion. For this simple problem, both are constant but depend on the tilt parameters which are time invariant here. Fig. \ref{Fig3} shows the evolving density and velocity of the underlying background fluid along with both phase and density contributions to vortex velocity. The left column of panels is associated with an untilted vortex, showing that it moves straight up. In contrast, the right column of panels shows that a tilted vortex will move at an oblique angle. The affect of tilt is identified by also plotting the density contribution to vortex velocity (red arrows) with $\Lambda = 1$ in Eq. \ref{v_final}. These predictions have only recently been experimentally verified~\cite{Andersen2021}.

To summarize, phase gradients correspond to the velocity of the underlying background fluid that may sweep a vortex along like a keeled boat drifting downriver without regard for its heading. A density gradient corresponds to a fluid pressure differential, generating a Magnus force~\cite{Magnus1987} that causes a vortex to move relative to the underlying fluid in a direction orthogonal to the gradient. But in such density gradients, the orientation of the keel of the boat is relevant as the relative motion between vessel and fluid could cause the boat to cut a path that need not be orthogonal to the density gradient. This is a general effect that could, in principle be observed in any experiment in which a Magnus force acts on an object that lacks azimuthal symmetry about its axis of rotation. 

%
%

\subsection{Trapped Linear Quantum Fluid} 

%
%
\begin{figure}[hptb]
	\begin{center}
		\includegraphics[width=0.48\textwidth]{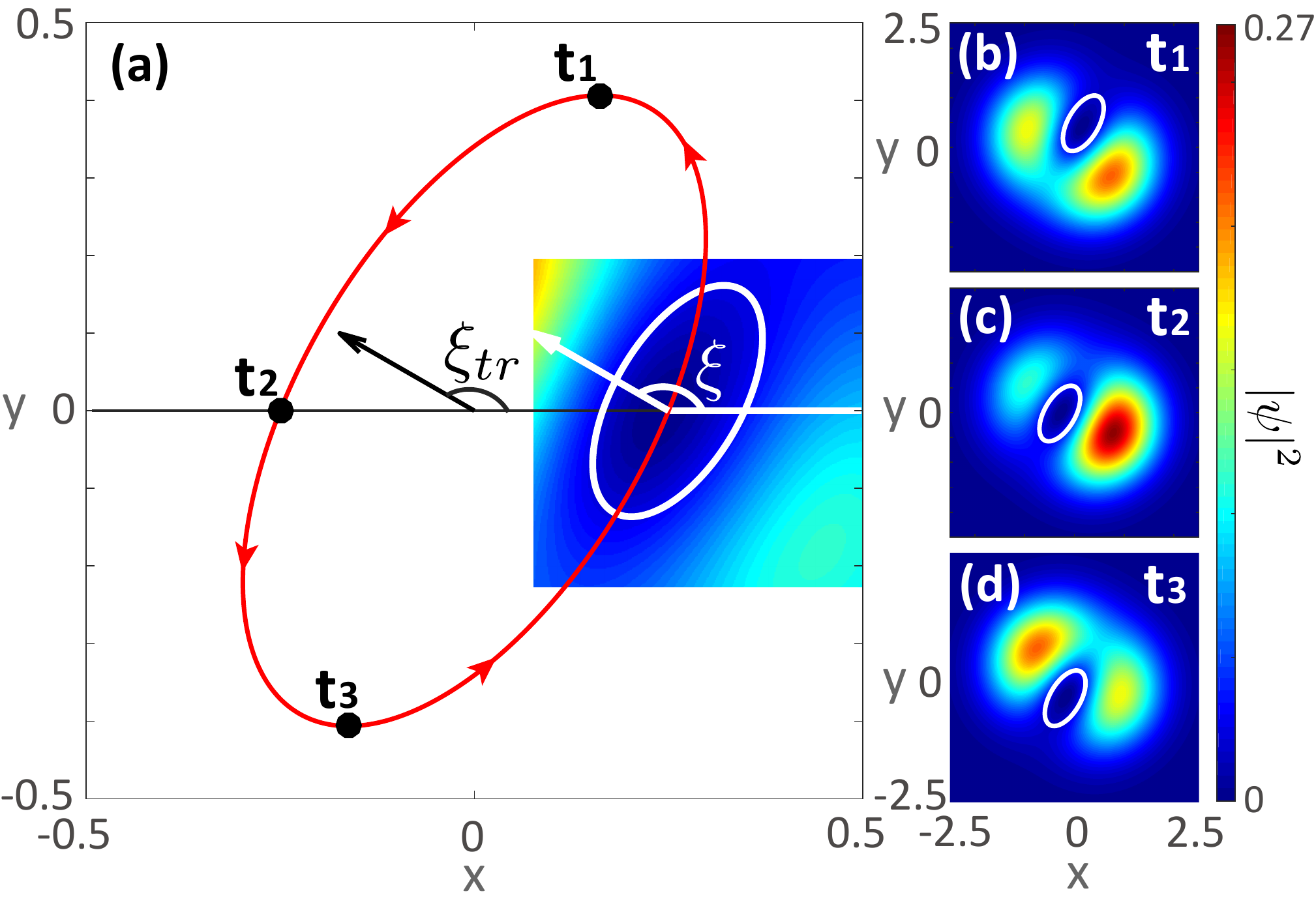}
	\end{center}
\caption{\emph{Dynamics of a tilted vortex in a linear quantum fluid ($\beta=0$).} The initial state is as shown in Fig.~\ref{Fig2} with $x_{0}=0.25$, $y_{0}=0$, $\xi_{0}=150^\circ$, and $\theta_{0}=60^\circ$. (a) The vortex trajectory analytically derived in Eq. \ref{Case1vel}. The vortex tilt $\xi(t)=\xi_{0}$ and polar lean $\theta(t)=\theta_{0}$ are stationary as the vortex moves, and its path is an ellipse characterized by these same tilt angles, i.e. $\xi_{tr}=\xi_{0}$ and $\theta_{tr}=\theta_{0}$. (b,c,d) The fluid density profiles for $t_{1}=\pi/2$, $t_{2}=\pi$, and $t_{3}=3\pi/2$, respectively, with specific contours highlighted to show that $\xi$ and $\theta$ are stationary.}
\label{Fig4}
\end{figure}
%

Next consider the motion of a vortex in a compressible, linear fluid moves that itself resides in a harmonic trap. The trap is accounted for with a harmonic potential, ${\cal V}=\frac{1}{2}(x^2 + y^2)$, in Eq. \ref{linearSE} and can be interpreted as the background fluid. The direct healing length of a linear-core vortex is infinity and, within our analytical framework, the typical size of the trap is 1, as shown in Fig. 4. Since this has no phase gradient, any vortex motion is solely due to an evolving gradient in the background fluid density. The background field can be applied to Eq.~\ref{v_final} to obtain the following prediction of vortex velocity:
\begin{eqnarray}\label{Case1vel}
	v_x &=& -x_0(\sin t + \cos t \cos \xi_0 \sin\theta_0 \sin\xi_0 \tan\theta_0) \cr
	v_y &=&  x_0 \cos t \sec\theta_0(\cos^2\xi_0 + \cos^2\theta_0\sin^2\xi_0) \, .
\end{eqnarray}
%
Here $\xi_0$ and $\theta_0$ describe the initial vortex tilt. The expressions for velocity can be easily integrated to obtain the vortex trajectory, $\{x_{v}\left(t\right), y_{v}\left(t\right)\}$. To more easily interpret this trajectory, consider a linear transformation that rigidly rotates the trajectory clockwise by the angle $\xi_{0}$:
\begin{equation}
\left[\begin{array}{cc}
\widetilde{x}_{v}\left(t\right) & \widetilde{y}_{v}\left(t\right)\end{array}\right]^{T}=R\left[\begin{array}{cc}
x_{v}\left(t\right) & y_{v}\left(t\right)\end{array}\right]^{T}.
\end{equation}
Here the rotation matrix is
\begin{equation}
R=\left(\begin{array}{cc}
\cos\xi_{0} & \sin\xi_{0}\\
-\sin\xi_{0} & \cos\xi_{0}
\end{array}\right).
\end{equation}
This results in the relationship
\begin{equation}
\frac{\left[\widetilde{x}_{v}\left(t\right)\right]^{2}}{\left(x_{0}\cos\theta_{0}\right)^{2}}+\frac{\left[\widetilde{y}_{v}\left(t\right)\right]^{2}}{x_{0}^{2}}=\sin^{2}\xi_{0}+\frac{\cos^{2}\xi_{0}}{\cos^{2}\theta_{0}},\label{eq:ellipse}
\end{equation}
which implies that the vortex trajectory is a fixed ellipse with azimuthal orientation $\xi_{tr}=\xi_{0}$ and aspect ratio $\cos\theta_{tr} = \cos\theta_{0}$. In addition, an explicit evaluation of the vortex tilt gives that the azimuthal angle and polar lean are both independent of time---i.e. $\xi\left(t\right)=\xi_{0}$ and
$\theta\left(t\right)=\theta_{0}$. This is shown in Fig.~\ref{Fig4} for a specific initial vortex tilt. If the vortex is initially untilted, the resulting circular orbit is consistent with earlier work \cite{one_circular_vortex1,one_circular_vortex2}. The direction of the trajectory is determined by the initial condition of the wave function. When $\xi_{0}$ is between $-90$ and $90$ degrees, as shown in Fig.~\ref{Fig4}, the trajectory is counter-clockwise; otherwise the trajectory is clockwise.

\section{Nonlinear Quantum Fluid} 
Attention is now turned to our primary focus, vortex dynamics in a trapped nonlinear fluid governed by the Gross-Pitaevskii equation (GPE),
\begin{equation}
	i\partial_{t}\psi = \left(-\frac{1}{2}\left(\partial_{xx}+\partial_{yy}\right)+\frac{1}{2}\left(x^{2}+y^{2}\right) + \beta\left|\psi\right|^{2}\right)\psi,
	\label{eq:GPE}
\end{equation}
with $\beta$ is the nonlinear interaction parameter.

This model for quantum fluids can be applied to 2D BEC by setting the atomic mass, the trap frequency, and $\hbar$ as characteristic units, regarding $t$ as the time and $\beta$ as the product of atom number and two-atom contact interaction strength~\cite{TOF}. It also captures the dynamics of paraxial optical fluid propagating through a nonlinear medium by treating the wave number and dielectric trap strength~\cite{Iadecola2016} as characteristic units, interpreting the Poynting vector axis as time, and identifying $\beta$ as the non-dimensional third-order susceptibility~\cite{Boyd2008}. 

Because analytical solutions do not exist for non-equilibrium vortex dynamics, the GPE is solved numerically~\cite{GPELab1,GPELab2}. Despite the nonlinearity, vortex velocity should still described by Eq. \ref{v_final}~\cite{Andersen2021}. That has not been previously verified and so is analyzed here.

\subsection{Weakly Nonlinear Quantum Fluid}
A tilted vortex is placed off-center in a weakly nonlinear quantum fluid ($\beta = 1$), and the GPE of Eq. \ref{eq:GPE} is numerically solved using an unconditionally stable relaxation pseudo-spectral scheme implemented on a $1024 \times 1024$ spectral grid~\cite{GPELab1,GPELab2}. Vortex position is subsequently obtained, at each time step, by using a root finder to determine the site at which both real and imaginary parts of the wave function are zero. The evolving vortex tilt is then obtained using a previously developed algebraic methodology~\cite{Andersen2021}. 

\begin{figure}[hptb]
	\begin{center}
		\includegraphics[width=0.48\textwidth]{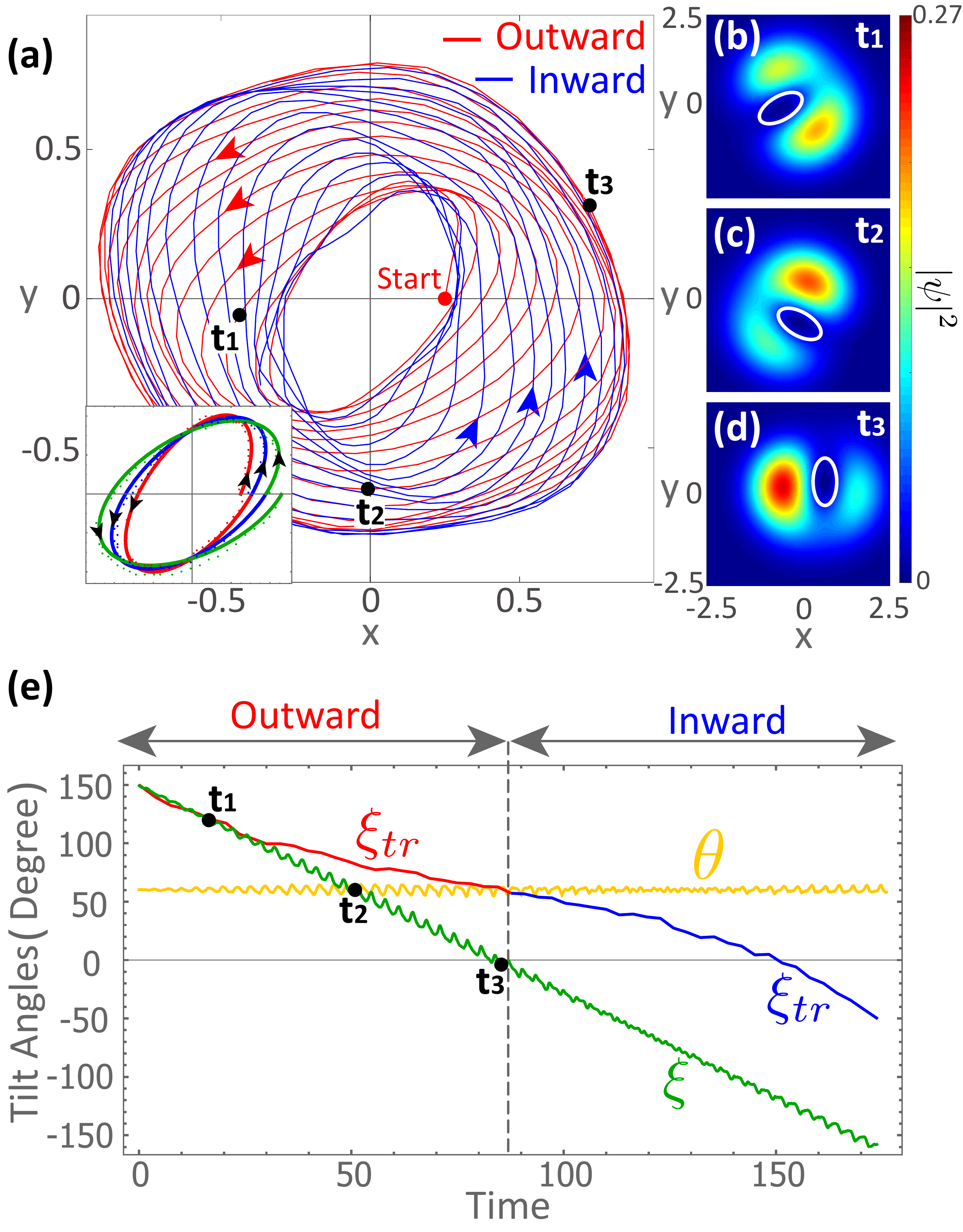}
	\end{center}
	\caption{\emph{Dynamics of a tilted vortex in a weakly nonlinear quantum fluid.} Nonlinear interaction $\beta=1$ and initial state is as shown in Fig.~\ref{Fig2} with $x_{0}=0.25$, $y_{0}=0$, $\xi_{0}=150^\circ$, and $\theta_{0}=60^\circ$. (a) The vortex trajectory for time period $t=0$ to $174$, in which the vortex spirals out (red) before spiraling back in again (blue). The first three circuits for the vortex spiralling out are plotted in the subfigure in (a). (b,c,d) The fluid density profiles for $t_{1}=16.6$, $t_{2}=51$, and $t_{3}=85.4$, respectively, corresponding to the black dots in (a,e), with specific contours highlighted to show the precession of vortex orientation. (e) Time evolution of azimuthal and polar angles of the vortex, $\xi\left(t\right)$ (green) and $\theta\left(t\right)$ (yellow), over one cycle of outward/inward spiraling shown along with evolution of azimuthal orientation of the trajectory, $\xi_{tr}\left(t\right)$ (red-blue). In the simulation, the spatial range is [-8 8] and the grid size is 1024 in x and y directions.}
	\label{Fig5}
\end{figure}

Strikingly, the nonlinear interaction generates a precession in the vortex azimuthal orientation, which is evident in the time slices of the fluid density shown in Fig.~\ref{Fig5}(b, c, d), where highlighted white contours give the vortex shape. As shown in Fig.~\ref{Fig5}(e), the vortex precesses clockwise as its azimuthal orientation $\xi(t)$ (green line) is a linearly decreasing function of time, while the polar lean $\theta(t)$ (yellow line) is constant over the entire simulation. A typical vortex trajectory is shown in Fig.~\ref{Fig5}(a), which exhibits several new features due to nonlinear interactions, and an animation of the dynamics is included in the Supplementary Material. The vortex now moves on a precessing elliptical trajectory, quantified by its evolving azimuthal orientation, $\xi_{tr}(t)$, plotted by the red (outward motion) and blue (inward motion) traces of Fig.~\ref{Fig5}(e). In addition, these ellipses grow and shrink, a new type of slow breathing mode. A comparison of $\xi(t)$ and $\xi_{tr}(t)$ indicates that the rate of precession of the trajectory mimics that of the vortex itself. When the vortex trajectory is in the smallest ellipse, the vortex and trajectory precess at the same rate just as in a linear fluid. This implies that structural character of the vortex can be quantified by observing the shape of the smallest trajectory. As the size of the ellipse grows, nonlinear effects cause the vortex to precess faster than the trajectory does. 

%
%
%

\begin{figure}
\includegraphics[scale=0.3]{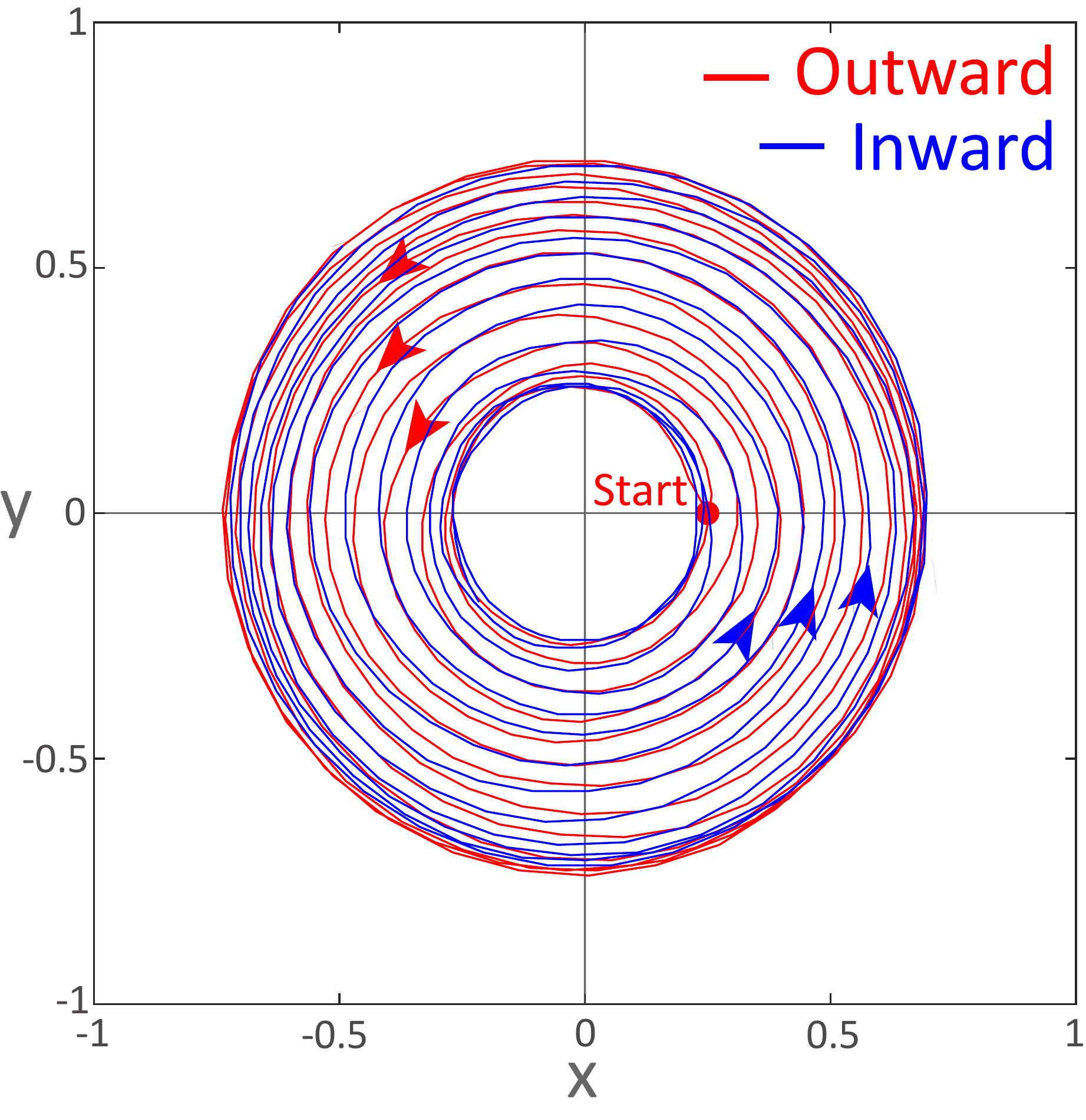}
\caption{The trajectory of an untilted, circular vortex for time period $t=0$
to $174$, in which the vortex spirals out (red) before spiraling back
in again (blue). The interaction strength is $\beta=1$ and the initial state
is described by $x_{0}=0.25$, $y_{0}=0$, and $\theta_{0}=0$. In the simulation, the spatial range is [-8 8] and the grid size is 1024 in x and y directions.}
\label{Fig6}
\end{figure}

The trajectory containing outward and inward spirals can also be observed
when the vortex is untilted. The circular vortex now moves in a circular trajectory, as plotted in Fig. \ref{Fig6}, where the red (blue) traces again the outward (inward) spirals. The weak nonlinearity
does not induce any meaningful tilt, so the vortex keeps its
circular shape with $\theta\left(t\right)=0$ during the time evolution.

%

\begin{figure}
\includegraphics[scale=0.29]{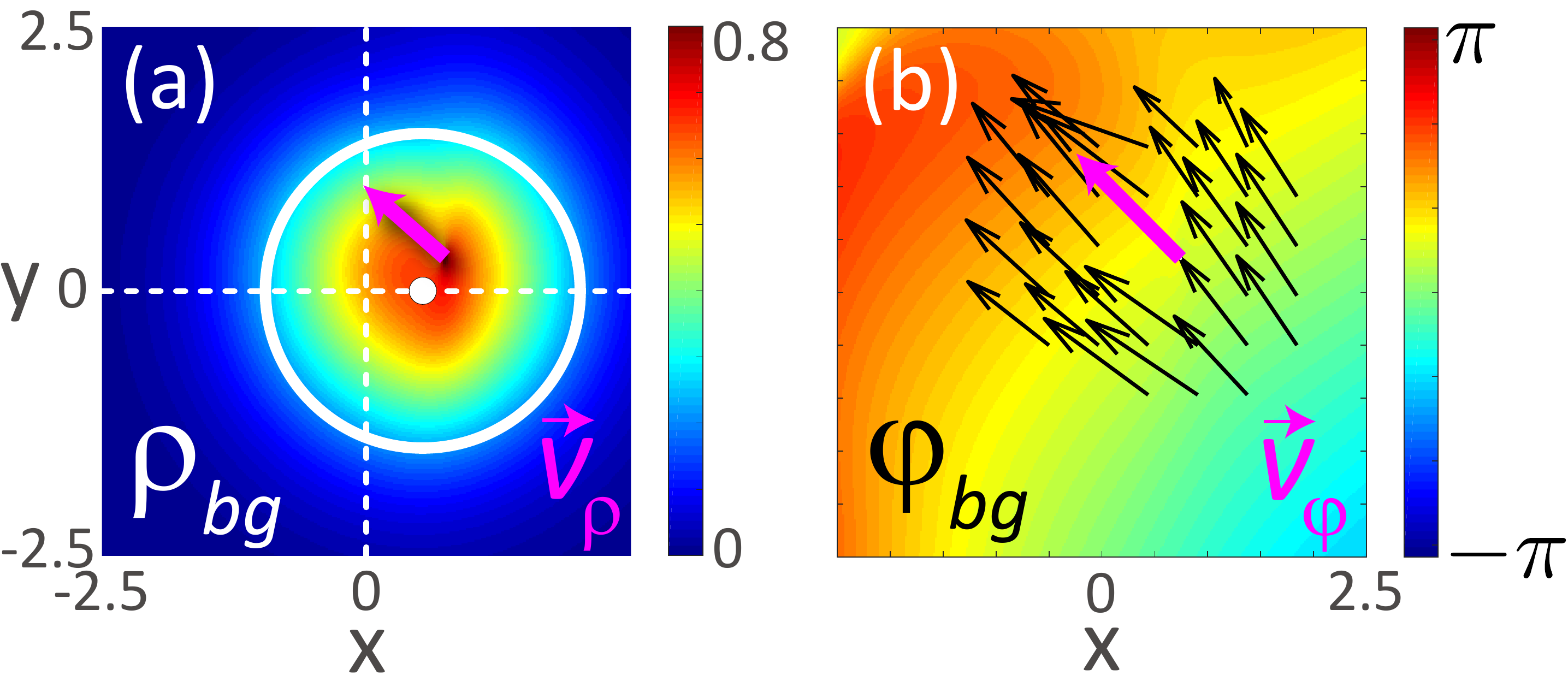}

\caption{(a) Background magnitude $\rho_{bg}$ and (b) background phase factor
$\varphi_{bg}$ at $t=85.4$ with magenta arrows denoting two times of velocities
$\vec{v}^{\rho}$ and $\vec{v}^{\varphi}$, respectively. In (a),
the white dot denotes the center of the background field and the white
circle denotes the shape of the approximated Gaussian profile. In
(b), the black arrows denote two times of $\vec{v}^{\varphi}$ calculated near
the vortex center. }

\label{Fig7}
\end{figure}

The vortex velocity relation of Eq. \ref{v_final} should predict even such complex vortex motion in a nonlinear quantum fluid. It requires that the gradients of the phase
$\varphi_{bg}$ and the amplitude $\rho_{bg}$ of the background field be evaluated at the vortex center. To obtain the background field, we divide out the field of
the vortex, $\psi_{bg}=\psi/\psi_{LC}$,
where $\psi$ is the total wave function of the quantum fluid obtained
from solving the Gross-Pitaevskii equation (GPE), and $\psi_{LC}$
is the wave function of a linear-core vortex given by
\begin{equation}\label{psiLC}
\psi_{LC}\left(x,y\right)=\left(x-x_{v}\right)a+\left(y-y_{v}\right)b.
\end{equation}
Here $a=-\cos\xi+i\cos\theta\sin\xi$ and $b=-\sin\xi-i\cos\theta\cos\xi$.
Note that the vortex-center coordinates, $x_{v}$ and $y_{v}$, and
the vortex tilt angles, $\xi$ and $\theta$, are all obtained from
the total wave function $\psi$. The tilted linear-core
vortex of Eq. \ref{psiLC} is then numerically divided out, and the resulting magnitude $\rho_{bg}$ and phase $\varphi_{bg}$ of the background field are shown in Fig. \ref{Fig7}.

The challenge in calculating the gradients of $\varphi_{bg}$ and
$\rho_{bg}$ at the vortex center is that the above quotient procedure above
causes a problematic numerical error for $\psi_{bg}$ near the vortex
center since the value of $\psi_{bg}$ at the vortex center is infinity.
To avoid this problem, we use separate approaches for estimating
$\vec{v}^{\rho}$ and $\vec{v}^{\varphi}$.

An estimate for $\vec{v}^{\rho}$ is obtained by approximating the background
amplitude as having a Gaussian profile:
\begin{equation}\label{eq:rhobg}
\rho_{bg}=e^{-\frac{1}{2}\left[\left(x-x_{b}\right)^{2}+\left(y-y_{b}\right)^{2}\right]}.
\end{equation}
Here \{$x_{b},y_{b}$\} is the numerically estimated center of the
background field, denoting by the white dot in Fig. \ref{Fig7}(a).
Note that the Gaussian center deviation from the origin is a manifestion of the
inward/outward spiral of the entire system. The circular shape of this
Gaussian profile is shown as a white circle in Fig. \ref{Fig7}(a),
in comparison to the real shape of the background amplitude. Since
the vortex position, \{$x_{v}$, $y_{v}$\}, and
background amplitude, $\rho_{bg}$ are known, it follows that
\begin{eqnarray}\label{eq:vrho}
v_{\rho x} &=&-y_{v}+y_{b} \nonumber \\
v_{\rho y} &=& x_{v}-x_{b},
\end{eqnarray}
where $v_{\rho x}$ and $v_{\rho y}$ are the $x$ and $y$ components
of $\vec{v}^{\rho}$, respectively. The magenta arrow in
Fig. \ref{Fig7}(a) denotes $\vec{v}^{\rho}$.

An estimate for $\vec{v}^{\varphi}$ is obtained by calculating and averaging the gradients
of $\varphi_{bg}$ for a set of locations in the neighborhood of the vortex center. These are indiated with black arrows in Fig. \ref{Fig7}(b) with their average giving the magenta arrow. 

%

\begin{figure}[hptb]
	\begin{center}
		\includegraphics[width=0.48\textwidth]{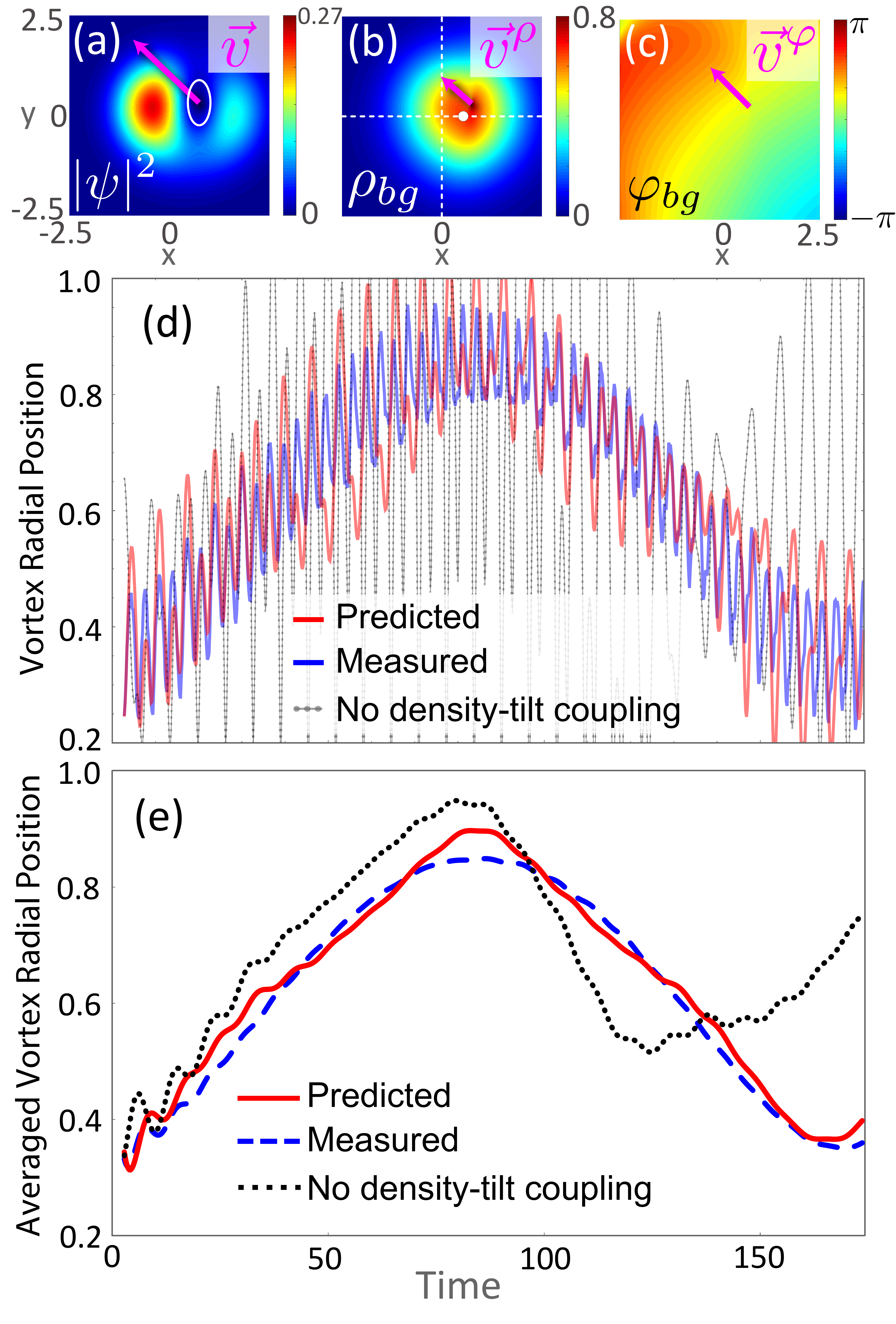}
	\end{center}
	\caption{\emph{Prediction versus simulation measurement of the radial position of a vortex.} The simulation of Fig.~\ref{Fig5} is used to assess the accuracy of the vortex velocity relation of Eq. \ref{v_final}. (a,b,c)  $t = t_{3}=85.4$, corresponding to Fig.~\ref{Fig5}(d): fluid density $|\psi|^2$, background magnitude $\rho_{bg}$, and background phase $\phi_{bg}$ with magenta arrows denoting two times of velocities $\vec{v}$, $\vec{v}^{\rho}$, and $\vec{v}^{\varphi}$, respectively, and a white dot denoting the center of background field. (d) Prediction of vortex radial position (red), obtained by integration of Eq. \ref{v_final} compared to the position measured (blue) using a root finder, as used in Fig.~\ref{Fig5}(a), to identify the evolving zero of the wave function. Also shown is the predicted radial position if coupling with tilt is not accounted for (black) by setting $\boldsymbol\Lambda = {\bf 1}$ in Eq. \ref{v_phi_v_rho}. (e) The result of the averaged vortex radial position, where the rapid cyclical oscillation in (d) is removed by low-pass filtering.}
	\label{Fig8}
\end{figure}

The methodology was applied to construct the vortex velocity components shown in Figs.~\ref{Fig8}(a, b, c) and to estimate the evolving vortex radial positions shown in Figs.~\ref{Fig8}(d, e).  Panel (b) shows that the extremum of the background density clearly deviates from the trap center, and the entire background field actually spirals cyclically in sync with the vortex. The background phase gradient shown in panel (c) is particularly interesting because it changes only slightly over the entire domain---i.e. it has a global character. All of these features are departures from what is observed for the linear media of Eq. \ref{Case1vel}, where the background field is on-center, without any phase gradient, and the vortex velocity is purely from $\vec{v}^{\rho}$. 

Gradients of the background fields (Figs.~\ref{Fig8}(b, c)) were used to calculate a radial velocity that was subsequently integrated to obtain the prediction for the radial evolution of the vortex (red) shown in Fig.~\ref{Fig8}(d). This compares favorably with the vortex position measured directly by the numerical simulation (blue). In both cases, low-pass filtering to remove rapid cyclical oscillations, as shown in Fig.~\ref{Fig8}(e), helps to more easily compare prediction with measurement over longer time scales. The background fluid density gradient is an essential contributor to the vortex velocity, and this is quantified in terms of the mean value of its contribution to the total vortex velocity of Eq. \ref{v_phi_v_rho}:
\begin{equation}
	\bigg<\frac{|\vec{v}^{\rho}|}{\sqrt{|\vec{v}^{\varphi}|^2 + |\vec{v}^{\rho}|^2}} \bigg> = 36\% .
\end{equation}
This ratio is calculated from the parameters for Figs.~\ref{Fig5} and \ref{Fig8}. The result is significantly different than for a harmonically trapped, linear fluid. There $v^{\varphi} = 0$ so the ratio is $100\%$. Panel (d) also shows curve (black) of what would be predicted for the radial position if the coupling between tilt and fluid density were not accounted for. This was produced by setting $\boldsymbol\Lambda = \bf 1$ in Eq. \ref{v_phi_v_rho}, and its poor prediction demonstrates how crucial it is to account for the newly identified coupling. 

In Fig.~\ref{Fig8}(d, e), the discrepancy between the predicted radial position (red) and the measured position (blue) is due to the approximations associated with both the numerical solution of the GPE and the methodology adopted to calculate background field gradients for the vortex velocity. The nonlinear GPE is numerically solved using a pseudo-spectral method implemented on a $1024\times1024$ spectral grid. The method is unconditionally stable, but errors are introduced by this finite grid size which accumulate as the vortex completes approximately 30 orbits. In addition, the finite grid size and large domain ($16\times 16$) imply that vortex position has a computational spatial uncertainty of $\pm 0.02$ in both x and y directions. 

In addition to the uncertainty associated with the simulator, the numerical implementation of the vortex velocity prediction of Eq. \ref{v_final} has its own sources of error. This is because it is populated with gradients in the background field evaluated at the center of the vortex, which is computationally problematic since the background field is the ratio of two fields that are both singular at the vortex center. As detailed above, the issue is addressed with two pragmatic idealizations: (i) that the background amplitude has a Gaussian profile; and (ii) that the background phase gradient at the vortex is equal to the average of its value in the surrounding neighborhood. These each contribute to an uncertainty in the vortex velocity prediction.

%
%

\subsection{Strongly Nonlinear Quantum Fluid}
The weak nonlinearity medium ($\beta=1$) is next replaced with a strongly nonlinear quantum fluid ($\beta=1000$). Instead of using the initial state of Eq. \ref{eq:psi0}, a linear-core vortex, the intial vortex profile is given a finite healing length compatible with the value of $\beta$ to make the vortex as stable as possible~\cite{Barenghibook}:
\begin{equation}
	\psi_{0}\left(x,y\right)=\phi_{gs}\left(x,y\right)\sqrt{1-\frac{l^{2}}{l^{2}+x'^{2}+y^{2}}}\frac{x'+iy}{\sqrt{x'^{2}+y^{2}}}.\label{eq:psi_s_circular}
\end{equation}
Here $x'=x-x_{0}$. The ground state of a nonlinear fluid without any vortex, $\phi_{gs}\left(x,y\right)$, is numerically obtained through imaginary time evolution, and the vortex healing length, $l$, is related to $\beta$ by
\begin{equation}
	l = \frac{1}{\sqrt{\beta}|\phi_{gs}\left(x_{0},0\right)|}.\label{eq:healing_l}
\end{equation}
To create a tilted-vortex initial condition with this healing length, Eq. \ref{eq:psi_s_circular} is modified by replacing $x'+ \imath y$ with $x'a +y b$, and replacing $x'^{2}+y^{2}$ with $|x'a+ y b|^{2}$. The new parameters, $a$ and $b$, are defined as
\begin{eqnarray}\label{abparams}
	a &=& -\cos\xi_{0}+i\cos\theta_{0}\sin\xi_{0} \nonumber \\
	b &=& -\sin\xi_{0}-i\cos\theta_{0}\cos\xi_{0}.
\end{eqnarray}
The initial state with an offset, tilted vortex in a strongly nonlinear quantum fluid is therefore given by 
\begin{equation}
	\psi_{0}\left(x,y\right)=\phi_{gs}\left(x,y\right)\sqrt{1-\frac{l^{2}}{l^{2}+|x'a+y b|^{2}}}\frac{x'a+y b}{|x'a+y b|}.\label{eq:psi_s_tilt}
\end{equation}

\begin{figure}[H]
	\begin{center}
		\includegraphics[width=0.48\textwidth]{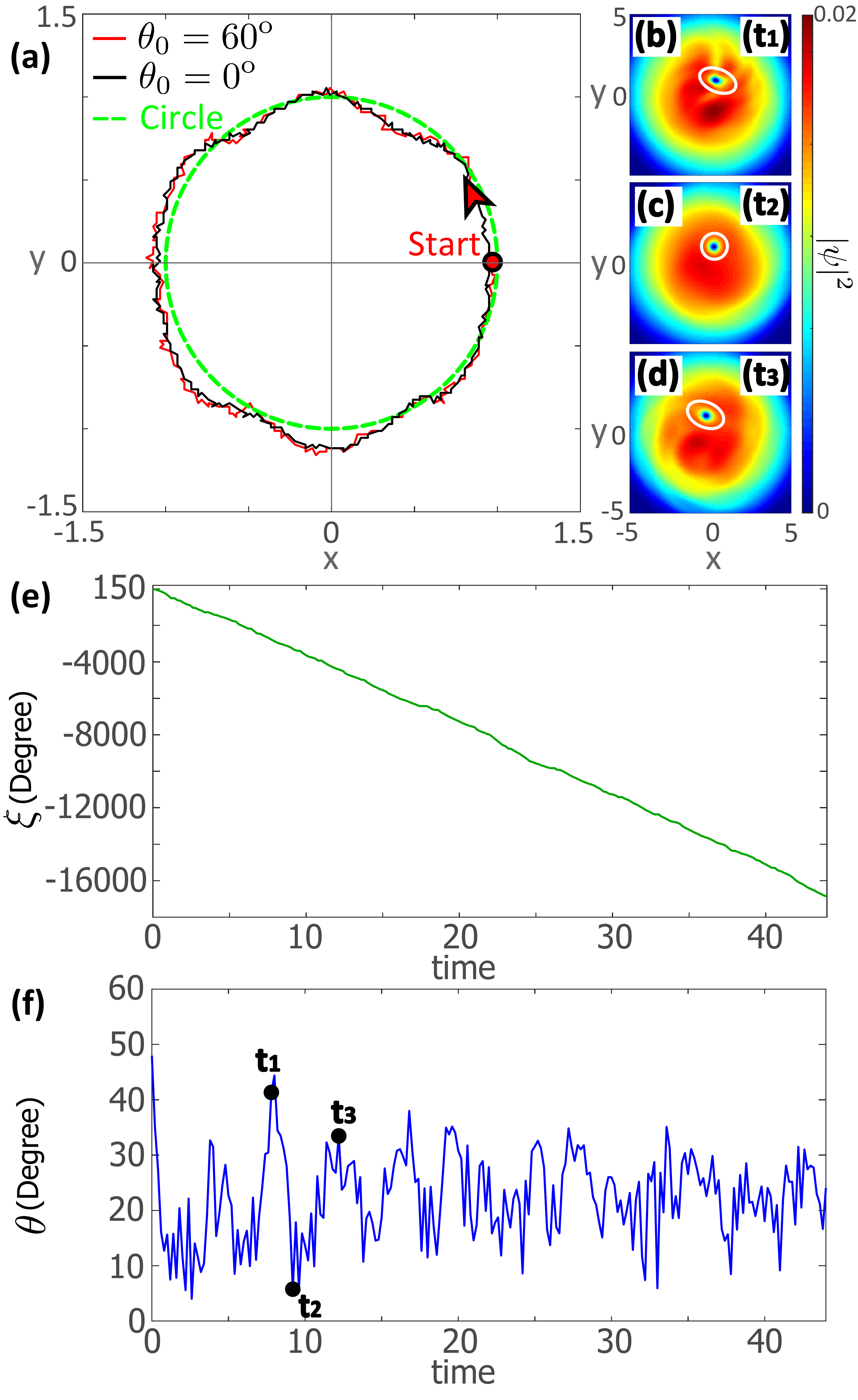}
	\end{center}
	\caption{\emph{Dynamics of an elliptical vortex in a quantum fluid with strong nonlinearity.} Nonlinear interaction $\beta=1000$ and vortex initial position $x_{0}=1$, $y_{0}=0$. (a) Vortex trajectories of the first loop for time period $t=0$ to $44$, for an elliptical vortex initialized with $\xi_{0}=150^\circ$ and $\theta_{0}=60^\circ$ in Eq. \ref{eq:psi_s_tilt} (red solid line), a circular vortex initialized with $\theta_{0}=0^\circ$ in Eq. \ref{eq:psi_s_circular} (black solid line), and a reference circular trajectory (green dashed line). (b, c, d) The fluid density profiles for $t_{1}=7.8$, $t_{2}=9.2$, and $t_{3}=12.2$, respectively, corresponding to the elliptical vortex in (a) and the black dots in (f), with specific contours highlighted to show the aspect ratio of vortex shape. (e) Time evolution of vortex azimuthal angle, $\xi\left(t\right)$, and (f) time evolution of vortex polar angle, $\theta\left(t\right)$, both corresponding to the evolution of the elliptical vortex in (a) for time period $t=0$ to $44$. In the simulation, the spatial range is [-16 16] and the grid size is 1024 in x and y directions.}
	\label{Fig9}
\end{figure}

Using the methodology introduced for the weakly nonlinear quantum fluid, the GPE of Eq. \ref{eq:GPE} is solved numerically with the results processed to obtain the evolving vortex position and tilt angles. This reveals two fundamental differences that can be seen by comparing Fig.~\ref{Fig9} with Fig.~\ref{Fig5}: (1) cyclic spiraling is now completely absent; and (2) the trajectory is now roughly circular instead of elliptical. The lack of spiral is due to the fact that the healing length is now extremely small, so the vortex is no longer influenced by condensate features on a larger length scale and, in particular, the position of the trap center. This property is inherited from the circular vortex case (black solid line), a setting in which it was previously observed by Polkinghorne et al.~\cite{Polkinghorne2021}. The circular character of the trajectory can be understood by noting, in Fig.~\ref{Fig9}(e), that precession in the vortex azimuthal angle, $\xi(t)$, is now extremely fast. In fact, the precession rate is approximately proportional to the interaction strength. For $\beta=1000$, the vortex rotates over 40 times for every degree of arc change in its orbit. As a result, the vortex ellipticity is averaged out, and the result is a circular motion as if the vortex itself was circular.

Another qualitative difference between weak and strongly nonlinear media is associated with the evolution of the polar lean, $\theta(t)$. For weak interactions, the polar lean is nearly constant, as shown in Fig.~\ref{Fig5}(e), but Fig.~\ref{Fig9}(f) shows that it undergoes substantial oscillations for strongly interacting fluids. This is also evident in the changes of vortex aspect ratio shown in the time slices of the fluid density, Fig.~\ref{Fig9}(b, c, d). Such temporal changes are actually the combined result of three separate contributions that are each amplified with increasing nonlinearity.

First consider a circular vortex at the center of a trapped, highly nonlinear condensate $\beta=1000$). The magnitude of variations in polar lean correlates inversely with grid size. For a $1024\times1024$ spectral grid, the polar lean varies randomly between $0^{\circ}$ and $20^{\circ}$, for a $2048\times2048$ spectral grid, the variations only range up to $10^{\circ}$, and for a $4096\times4096$ spectral grid the random variation in measured polar lean has a range of only $5^{\circ}$. Such noisy character of polar lean can therefore be attributed to numerical error, most likely the uncertainty in vortex position.

When the vortex at the trap center, an elliptical vortex exhibits an additional oscillation in the polar lean that increases with the degree of nonlinearity. This is due to an anisotropic squeezing force applied to the elliptical vortex by the circular background quantum fluid. The vortex is squeezed more heavily along the direction of the longer axis than that of the shorter axis. This effect is analogous to the evolution of an elliptical Bose-Einstein condensate in a circular trap, where the shape of the condensate is periodically squeezed along the longer axis and expanded along the shorter axis. The result is that the polar lean rapidly decreases and then oscillates about a lower average value. For a fluid with $\beta=1000$ and an initial polar lean of $60^{\circ}$, the average polar lean reduces to $25^{\circ}$ with an oscillation of approximately $\pm 20^{\circ}$.

Finally, consider the case of an off-center tilted vortex, as shown in Fig. \ref{Fig9}(f). Now the non-uniform background field introduces a further density gradient in the fluid around the vortex which causes an additional squeezing force applied to the elliptical vortex. The result is that an off-center vortex once again evolves into an less tilted aspect ratio, but now the subsequent oscillations are only on the order of the random error observed for the centered case (Fig. \ref{Fig9}(f)).

%
%

\subsection{Qualitative Explanation for Vortex Precession in a Nonlinear Quantum Fluid}

\begin{figure}[H]
	\begin{center}
		\includegraphics[width=0.48\textwidth]{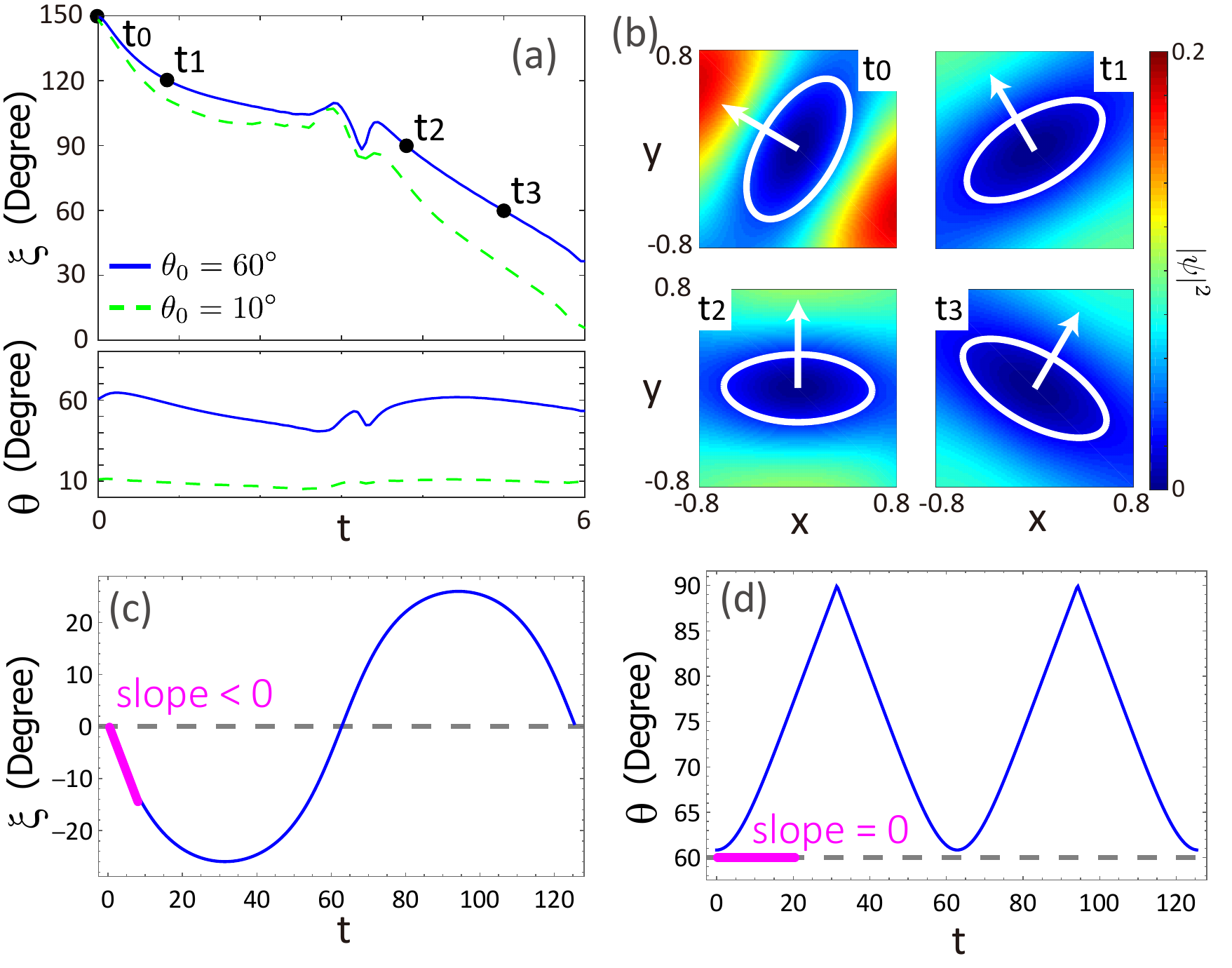}
	\end{center}
\caption{\emph {Dynamics of an on-center elliptical vortex in a nonlinear quantum fluid.} Interaction strength $\beta=20$ and initial azimuthal angle $\xi_{0}=150^\circ$. 
(a) Time evolution of azimuthal angle $\xi(t)$ and polar angle $\theta(t)$ with trap frequency $\omega=1$, where the numerical result is for $\theta_{0}=10^\circ$ and $60^\circ$, and the analytical estimate is with $\left\langle h\right\rangle =0.0417$ valid for small $\theta$; (b) Time sequence of fluid densities corresponding to black points in panel (a), where white lines show density contours and white arrows indicate $\xi(t)$; (c, d) Time evolution of vortex tilt for an elliptical trap idealization with $\gamma=0.95$.}
	\label{Fig10}
\end{figure}

For both weakly (Fig.~\ref{Fig5}(e)) and strongly (Fig. \ref{Fig9}(e)) nonlinear media, vortex precession is characterized by a linear evolution of the azimuthal orientation, $\xi$ with time.  Consistent with these results, Fig. \ref{Fig10}(a) shows that azimuthal orientation, $\xi(t)$, decreases at a relatively constant rate for an intermediate value of nonlinear interaction strength, $\beta = 20$. To focus squarely on the precession, the vortex has been placed at the center of the trap so that its orientation precesses but its center stays fixed. This precession is only weakly dependent on the initial vortex lean $\theta_0$, and the polar lean itself, $\theta(t)$, is relatively constant with time. These features are also evident in the time slices of the field density, shown in Fig. \ref{Fig10}(b), where highlighted white contours show the vortex shape and associated white arrows denote the evolving azimuthal angle.

%

Both of these characteristics can be explained with a simple idealization in which the nonlinear term in GPE is absorbed into the trap strength using
\begin{equation}\label{eq:mtl_1}
\beta |\psi|^2 \approx \beta |\psi_{\rm LC, init}|^2 .
\end{equation}
Vortex evolution is then governed by a linear system with an elliptical trap,
\begin{equation}\label{eq:Hellip}
{\rm H_{ellip}}=-\frac{\hbar^{2}}{2m}\left(\partial_{xx} + \partial_{yy}\right)+\frac{1}{2}m\omega^{2}\left(x^{2}+\gamma^2 y^{2}\right),
\end{equation}
where $\gamma$ and $\omega$ are functions of $\beta$ and the initial vortex tilt angles. The associated eigenmodes are products of Hermite polynomials, and the initial condition is reasonably approximated as the sum of the lowest pair of excited modes, $\left|01\right>$ and $\left|10\right>$. The evolving field is then of the form
\begin{equation}\label{eq:bobbing}
\left|\psi(t)\right> = {\it N}(\left|01\right> + \alpha e^{\imath (\varepsilon_{10}-\varepsilon_{01})t}\left|10\right>).
\end{equation}
The ellipticity-induced difference in mode energies, $\varepsilon_{10}-\varepsilon_{01}$, results in a beating phenomenon that is seen as the vortex lean bobbing up and down as shown in Fig. \ref{Fig10}(d). Likewise, the ellipticity-induced weighting coefficient, $\alpha$, imbalances what would otherwise be an azimuthal standing mode, and the vortex orientation oscillates back and forth. The magenta lines in panel Figs. \ref{Fig10}(c,d) emphasize that the rate of change of vortex orientation, $\dot \xi$, is initially constant and negative, while the rate of change of polar lean, $\dot \theta$, is initially zero. The nonlinear interaction in the GPE, though, amounts to a self-trap that rotates with the vortex, implying that the idealized tilt dynamics hold for all times in a Zeno-like manner. This explains the trends observed in the numerical results of Fig. \ref{Fig10}(a).

\FloatBarrier

\section{Conclusion}

We have shown that it is possible to quantitatively predict vortex trajectories in quantum fluids by accounting for the coupling between vortex tilt and the background quantum state. This coupling is negligible in regimes that are well-approximated as incompressible, but they are particularly relevant when the vortex healing length is on the order of vortex separation in few-body systems. In the absence of nonlinear interactions, the background field depends only on the trap, and isolated tilted vortices move in an elliptical path that is self-similar to their own projection. Nonlinear effects allow for richer dynamics, though, since the vortex can now contribute to its own background field. The effect amounts to the vortex being able to influence its own motion and tilt. Significantly, vortex tilt and its rate of precession are mimicked in the trajectory observed, allowing these important local features to be measured with relative ease. This capability, in turn, is expected to be useful in developing on-the-fly manipulation of trap strength and atomic interaction as a means of controlling few-body vortex interactions such as nucleation, annihilation, scattering, and braiding. 

\section{Acknowledgement}

The authors acknowledge useful discussions with Jasmine Andersen and Drew Voitiv. We are grateful to the W. M. Keck Foundation and the National Science Foundation (DMR 1553905) for supporting this research.


\end{document}